\documentclass{svproc}
\usepackage{cite}
\usepackage[colorlinks=true,linkcolor=red,
            urlcolor=gray,
            citecolor=blue]{hyperref}
\usepackage{graphicx} 
\usepackage{latexsym}
\usepackage{tensor}
\usepackage{orcidlink}
\def\tns{\tensor}
\newcommand{\be}{\begin{equation}}
\newcommand{\ee}{\end{equation}}
\newcommand{\bear}{\begin{array}}
\newcommand{\eear}{\end{array}}
\newcommand{\ba}{\begin{eqnarray}}
\newcommand{\ea}{\end{eqnarray}}

\begin{document}
\mainmatter              
\title{Cosmic inflation in metric-affine gravity}
\titlerunning{Cosmic inflation in metric-affine gravity}  
%
\author{Ioannis~D.~Gialamas\orcidlink{0000-0002-2957-52765}\inst{1} \and Kyriakos~Tamvakis\orcidlink{0009-0007-7953-9816}\inst{2}}
\authorrunning{I.~D.~Gialamas and K.~Tamvakis} 
%

%
\institute{Laboratory of High Energy and Computational Physics, 
National Institute of Chemical Physics and Biophysics, R{\"a}vala pst.~10, Tallinn, 10143, Estonia\\[0.1cm]
\and
Physics Department, University of Ioannina, 45110, Ioannina, Greece}

\maketitle              

\begin{abstract}
In the context of metric-affine gravity theories, where the metric and connection are independent, we examine actions involving quadratic terms in the Ricci scalar curvature and the Holst invariant. These actions are non-minimally coupled to a scalar field. We explore the behavior of the corresponding effective metric theory, which includes an extra dynamic pseudoscalar degree of freedom.  Detailed analysis of inflationary predictions reveals compliance with recent observations across various parameters, potentially allowing a higher tensor-to-scalar ratio. The spectral index's direction of change varies based on parameter positioning.
\end{abstract}
\section{Introduction}
The present framework of modern cosmology consists of classical General Relativity (GR) as a theory of gravitation and Quantum Field Theory (QFT) as the theory of matter. A common working assumption is that the quantum aspects of gravitation can be ignored for energies below the Planck energy of $10^{19}\, {\rm GeV}$ and, therefore, gravity can be treated classically. In contrast, the full quantum character of particle interactions is considered within QFT. The quantum interactions of the matter fields coupled to the classical gravitational field introduce modifications to the standard GR action with cosmological implications. Such are non-minimal couplings of the inflaton field or higher power terms of the Ricci curvature in models of cosmological inflation. The metric-affine formulation of gravity, where the metric and the connection are independent variables, although equivalent to the standard (metric) GR in the case of the Einstein-Hilbert action, leads to different predictions when the above corrections are included.

\section{Metric vs metric-affine formulation of gravity}
The General Relativity Principle states that all laws of physics should be invariant under general coordinate transformations. To implement such a principle we need to introduce a metric $g_{ \mu\nu}$, which has to transform as
\begin{equation}
g'_{ \alpha\beta}(x')=\left(\frac{\partial x^{ \mu}}{\partial{x'}^{ \alpha}}\right)\left(\frac{\partial x^{ \nu}}{\partial{x'}^{ \beta}}\right)g_{ \mu\nu}(x)\,,    
\end{equation} 
as well as a Connection $\Gamma_{ \mu\,\,\,\nu}^{ \,\,\,\rho}$ in order to define covariant derivatives of tensors. In the standard metric formulation of gravity the connection is not an independent quantity but it is given by the Levi-Civita relation as
\begin{equation}
\left.\Gamma_{ \mu\,\,\,\nu}^{ \,\,\,\rho}\right|_{LC}=\frac{1}{2}g^{ \rho\sigma}\left(\partial_{ \mu}g_{ \rho\nu}+\partial_{ \nu}g_{\mu\sigma}-\partial_{ \sigma}g_{ \mu\nu}\right)\,.\label{LC}
\end{equation} 
In contrast, in the so-called metric-affine theories of gravity the connection is an independent variable not related to the metric through~(\ref{LC}). Note that $D_{ \mu}g_{ \nu\rho}\big|_{LC}=0$ (metricity), while $D_{ \mu}g_{\nu\rho}\neq 0$ in general for a metric-affine theory.

\subsection{The distortion tensor}
The difference between the independent connection of a metric-affine theory and the corresponding Levi-Civita one is a tensor called the distortion tensor
\begin{equation} 
C_{ \mu\,\,\,\nu}^{\,\,\,\rho}=\Gamma_{ \mu\,\,\,\nu}^{\,\,\, \rho}\,-\left.\Gamma_{ \mu\,\,\,\nu}^{ \,\,\,\rho}\right|_{LC}\,.
\end{equation} 
The distortion tensor vanishes for metric theories. The Torsion is given by 
$T_{ \mu\,\,\,\nu}^{\,\,\,\rho}=C_{ \mu\,\,\,\nu}^{\,\,\,\rho}-C_{ \nu\,\,\,\mu}^{\,\,\,\rho}=2C_{[\mu\,\,\,\nu]}^{\,\,\,\,\rho}=\Gamma_{ \mu\,\,\,\nu}^{\,\,\,\rho}-\Gamma_{ \nu\,\,\,\mu}^{\,\,\,\rho}$, while the curvature tensor of a metric-affine theory is defined as
\begin{equation} 
{\cal{R}}_{ \mu\nu\,\,\,\sigma}^{\,\,\,\,\,\,\rho}=\partial_{ \mu}\tns{\Gamma}{_\nu^\rho_\sigma}-\partial_{ \nu}\tns{\Gamma}{_\mu^\rho_\sigma}+\tns{\Gamma}{_\mu^\rho_\lambda}\tns{\Gamma}{_\nu^\lambda_\sigma}-\tns{\Gamma}{_\nu^\rho_\lambda}\tns{\Gamma}{_\mu^\lambda_\sigma}\,.
\end{equation} 

The following two scalars, linear in the Riemann tensor, can be defined as
\begin{eqnarray} 
\cal{R}&=&\tns{\cal{R}}{_\mu_\nu^\rho_\sigma}\delta_{\rho}^{ \mu}g^{\nu\sigma}=\tns{\cal{R}}{_\mu_\nu^\mu^\nu}  \,,
\\ \tilde{\cal{R}}&=&(-g)^{-1/2}\tns{\epsilon}{^\mu^\nu_\rho^\sigma} \tns{\cal{R}}{_\mu_\nu^\rho_\sigma}=(-g)^{-1/2}\tns{\epsilon}{^\mu^\nu^\rho^\sigma}\tns{\cal{R}}{_\mu_\nu_\rho_\sigma}\,.
\end{eqnarray} 

The first corresponds to the usual Ricci scalar, while $\tilde{\cal{R}}$ is the so-called Holst invariant, which vanishes identically in a metric theory due to the symmetry in the lower indices of the Levi-Civita connection. The Holst invariant commonly appears in Loop Quantum
Gravity~\cite{Rovelli:1997yv} and has been studied extensively in inflationary cosmology~\cite{Langvik:2020nrs,Shaposhnikov:2020gts,Piani:2022gon,Rigouzzo:2022yan,Pradisi:2022nmh,Salvio:2022suk,Gialamas:2022xtt,Gialamas:2023emn,Gialamas:2024iyu}.

The following expressions of the curvature scalars can be written in terms of the distortion
\begin{eqnarray} 
{\label{FORM-C1}}
{\cal{R}}&=& R+D_{ \mu}C_{ \nu}^{\,\,\,\mu\nu}-D_{ \nu}C_{ \mu}^{\,\,\,\mu\nu}+C_{ \mu\,\,\,\,\lambda}^{\,\,\,\mu}C_{ \nu}^{\,\,\lambda\nu}-C_{ \nu\,\,\,\lambda}^{\,\,\,\mu}C_{ \mu}^{\,\,\,\lambda\nu}
\,,\\
\tilde{\cal{R}}&=&2(-g)^{-1/2}\epsilon^{ \mu\nu\rho\sigma}\left(D_{ \mu}C_{ \nu\rho\sigma}+C_{ \mu\rho\lambda}C_{ \nu\,\,\sigma}^{\,\,\lambda}\right)\,,
{\label{FORM-C}}
\end{eqnarray}
where $R=R[g]$ is the standard metric Ricci scalar and the covariant derivatives are taken with respect to $\left.\Gamma_{ \mu\,\,\nu}^{\,\,\rho}\right|_{LC}$.

The metric-affine version of the Einstein-Hilbert action is
\begin{eqnarray}
{\cal{S}}_{EH}&=&\frac{1}{2}\int{\rm d}^4x\sqrt{-g}{\cal{R}} \nonumber
\\ &=& \frac{1}{2}\int {\rm d}^4x\sqrt{-g}\left\{R+D_{ \mu}C_{ \nu}^{\,\,\,\mu\nu}-D_{ \nu}C_{ \mu}^{\,\,\,\mu\nu}+C_{ \mu\,\,\,\,\lambda}^{\,\,\,\mu}C_{ \nu}^{\,\,\lambda\nu}-C_{ \nu\,\,\,\lambda}^{\,\,\,\mu}C_{ \mu}^{\,\,\,\lambda\nu}\right\}\,.
\end{eqnarray}
Variation with respect to the distortion gives 
\begin{equation}
\frac{\delta {\cal{S}}}{\delta C}=0\,\,\,\Longrightarrow\,\delta_{ \beta}^{ \alpha}C_{ \nu\gamma}^{\,\,\,\,\,\,\nu}+\delta_{ \gamma}^{ \alpha}C_{ \nu\,\,\,\,\beta}^{ \,\,\,\,\nu}-C_{ \beta\gamma}^{\,\,\,\,\,\,\,\alpha}-C_{ \gamma\,\,\,\beta}^{\,\,\,\alpha}=0\,.
\end{equation}
This equation possesses the general solution $C_{\mu\nu\rho}=U_{\mu}g_{\nu\rho},$ in terms of the arbitrary vector $U_{ \mu}$. Upon inserting this solution into the action, the terms that rely on $C$ disappear. Consequently, $ {\cal{S}}_{EH}$ is equivalent to the conventional metric GR. Nevertheless, this equivalence does not extend to quadratic actions.

\section{Non-minimal coupling to scalars}
One can derive the metric-equivalent of any metric-affine theory based on an action, where gravity couples to a scalar field,
\begin{equation}
{\cal{S}}=\int\,{\rm d}^4x\,\sqrt{-g}\left\{\frac{1}{2}\Omega^2(\phi){\cal{R}}\,+\,{\cal{L}}(\phi,g_{ \mu\nu},\,\partial_{ \mu}\phi)\right\}\,.\
\end{equation}
Note that any $F({\cal{R}})$ theory can also be set in this form, i.e.
\begin{equation}
 {\cal{S}}=\int{\rm d}^4x\sqrt{-g}\left\{\frac{1}{2}F'(\chi){\cal{R}}-V(\chi)\right\}\,,\quad\textit{where}\quad V(\chi)=\frac{1}{2}\left(\chi F'(\chi)-F(\chi)\right)\,, 
\end{equation}
in terms of the auxiliary scalar $\chi$.

Substituting the expression of ${\cal{R}}$ in terms of the distortion, we obtain
\begin{eqnarray}
{\cal{S}}=\int\,{\rm d}^4x\,\sqrt{-g}&&\left\{\frac{1}{2}\Omega^2(\phi)R(g)+\frac{1}{2}\Omega^2(\phi)\left(D_{ \mu}C_{ \nu}^{\,\,\,\mu\nu}-D_{ \nu}C_{ \mu}^{\,\,\,\mu\nu}\,\right.\right. \nonumber
\\ &&
\left.\left.+\,C_{ \mu\,\,\,\,\lambda}^{\,\,\,\mu}C_{ \nu}^{\,\,\lambda\nu}-C_{ \nu\,\,\,\lambda}^{\,\,\,\mu}C_{ \mu}^{\,\,\,\lambda\nu}\right) 
\,+\,{\cal{L}}(\phi,g_{ \mu\nu},\,\partial_{ \mu}\phi)\right\}\,.
\end{eqnarray}
Variation with respect to the distortion gives us the equation
\begin{equation}
\Omega^2\left(\delta_{ \beta}^{ \alpha}C_{ \nu\gamma}^{\,\,\,\,\,\,\nu}+\delta_{ \gamma}^{ \alpha}C_{ \nu\,\,\,\beta}^{\,\,\nu}-C_{ \beta\gamma}^{\,\,\,\,\,\,\,\alpha}-C_{ \gamma\,\,\,\beta}^{\,\,\alpha}\right)=\delta_{ \gamma}^{ \alpha}\partial_{ \beta}\Omega^2-\delta_{ \beta}^{ \alpha}\partial_{ \gamma}\Omega^2\,,
\end{equation}
with a solution (up to a term  $U_{\mu}\,g_{\nu\rho}\,$ of an arbitrary vector $U_{ \mu}$)
\begin{equation}
C_{ \mu\,\,\nu}^{\,\,\,\rho}=\frac{1}{2}\left(g^{ \rho}_{ \mu}\partial_{ \nu}\ln\Omega^2+g^{ \rho}_{ \nu}\partial_{ \mu}\ln\Omega^2-g_{ \mu\nu}\partial^{ \rho}\ln\Omega^2\right)\,.
\end{equation}
Substituting the distortion $C$ back into the action we obtain
\begin{equation}
{\cal{S}}=\int\,{\rm d}^4x\,\sqrt{-g}\left\{\frac{1}{2}\Omega^2(\phi){{R}}(g)\,+\,\frac{3}{4}\frac{(\nabla\Omega^2)^2}{\Omega^2}\,+\,{\cal{L}}(\phi,g_{ \mu\nu},\,\partial_{ \mu}\phi)\right\}\,.
\end{equation}
This is a metric theory and the appearing connection is the Levi-Civita one. Note that the extra term has the form of the extra kinetic term that appears when we Weyl-rescale the metric theory to the Einstein frame, albeit with an opposite sign. The inequivalence of the two formulations rests on this term. Only in the case of the Einstein-Hilbert action this term vanishes and the two formulations are equivalent.

\section{Quadratic metric-affine theories}

Consider the following metric-affine generalization of the Starobinsky model~\cite{Starobinsky:1980te}
\be {\cal{S}}=\int{\rm d}^4x\sqrt{-g}\left\{\frac{1}{2}\alpha{\cal{R}}+\frac{1}{2}\beta\tilde{\cal{R}}+\frac{1}{4}\gamma{\cal{R}}^2+\frac{1}{4}\delta\tilde{\cal{R}}^2\,\right\}\,,\ee 
where ${\cal{R}}$ is the Ricci scalar curvature and $\tilde{\cal{R}}$ is the Holst invariant. This is a general quadratic action of these scalars. Nevertheless, Palatini inflationary models (i.e. metric-affine models with zero torsion) of a scalar field in the presence of an $R^2$ term have been studied with predictions compatible with existing cosmological data~\cite{Enckell:2018hmo,Antoniadis:2018ywb,Bombacigno:2018tyw,Antoniadis:2018yfq,Edery:2019txq,Tenkanen:2019wsd,Gialamas:2019nly,Lloyd-Stubbs:2020pvx,Antoniadis:2020dfq,Ghilencea:2020piz,Das:2020kff,Tang:2020ovf,Gialamas:2020snr,Iosifidis:2018zjj,Iosifidis:2020dck,Dimopoulos:2020pas,Gialamas:2020vto,Karam:2021sno,Lykkas:2021vax,Gomez:2021roj,Gialamas:2021enw,Antoniadis:2021axu,Annala:2021zdt,Dioguardi:2021fmr,Gialamas:2021rpr,AlHallak:2022gbv,Dimopoulos:2022tvn,Dioguardi:2022oqu,Panda:2022esd,Dimopoulos:2022rdp,Durrer:2022emo,Antoniadis:2022cqh,Lahanas:2022mng,Gialamas:2022gxv,Panda:2022can,Gialamas:2023flv,Gialamas:2023aim,Dioguardi:2023jwa,Gialamas:2024jeb}.

In what follows we shall use Planck-mass units taking $\alpha=1$. An equivalent way to express the action is in terms of the auxiliary scalars $\chi$ and $\zeta$ as
\be {\cal{S}}=\int{\rm d}^4x\sqrt{-g}\left\{\frac{1}{2}(1+\gamma\chi){\cal{R}}+\frac{1}{2}(\beta+\delta\zeta)\tilde{\cal{R}}-\frac{1}{4}\left(\gamma\chi^2+\delta\zeta^2\right)\,\right\}\,.\ee
Next, we may use the expressions of ${\cal{R}}$ and $\tilde{\cal{R}}$ in terms of the distortion $C$, given in ({\ref{FORM-C1}),(\ref{FORM-C}}), and obtain the equivalent metric theory. Variation with respect to $C_{ \alpha}^{\,\,\beta\gamma}$ gives\footnote{See~\cite{Iosifidis:2021bad} for a general solution of this type of equations and~\cite{Gialamas:2022xtt,Gialamas:2023emn} for more details on the solution adopted here. }
\be \frac{\Omega^2}{2}\left(\delta_{ \beta}^{ \alpha}C_{ \nu\gamma}^{\,\,\,\,\,\,\nu}+\delta_{ \gamma}^{ \alpha}C_{ \nu\,\,\,\beta}^{\,\,\nu}-C_{ \beta\gamma}^{\,\,\,\,\,\,\,\alpha}-C_{ \gamma\,\,\,\beta}^{\,\,\alpha}\right)-\frac{\overline{\Omega}^2}{\sqrt{-g}}\left(\epsilon^{ \mu\alpha\sigma}_{\,\,\,\,\,\,\,\,\,\,\beta}C_{ \mu\gamma\sigma}+\epsilon^{ \mu\alpha\sigma}_{\,\,\,\,\,\,\,\,\,\,\gamma}C_{ \mu\sigma\beta}\right)=J_{ \beta\,\,\,\gamma}^{\,\,\alpha}\,,\ee
where
\be
J_{ \beta\,\,\,\gamma}^{\,\,\alpha}=\frac{1}{2}\delta_{ \gamma}^{ \alpha}\partial_{ \beta}\Omega^2-\frac{1}{2}\delta_{ \beta}^{ \alpha}\partial_{ \gamma}\Omega^2+\frac{\epsilon^{ \mu\alpha}_{\,\,\,\,\,\,\,\beta\gamma}}{\sqrt{-g}}\partial_{ \mu}\overline{\Omega}^2\,,
\ee
and $\Omega^2\equiv 1+\gamma\chi,\,\,\,\,\,\,\overline{\Omega}^2=\beta+\delta\zeta
$. Note that in the previous equation the RHS is antisymmetric in the lower indices, i.e. $J_{ \beta\,\,\,\gamma}^{\,\,\alpha}=-J_{ \gamma\,\,\,\beta}^{\,\,\alpha}$. Then, we obtain the following solution
$$ C_{ \mu\nu\rho}=\frac{g_{ \mu\nu}}{2\Delta}\left(\Omega^2\partial_{ \rho}\Omega^2+4\overline{\Omega}^2\partial_{ \rho}\overline{\Omega}^2\right)-\frac{g_{ \mu\rho}}{2\Delta}\left(\Omega^2\partial_{ \nu}\Omega^2+4\overline{\Omega}^2\partial_{ \nu}\overline{\Omega}^2\right)$$
\be+\frac{\epsilon_{ \mu\nu\rho\sigma}}{\Delta\sqrt{-g}}\left(\Omega^2\partial^{ \sigma}\overline{\Omega}^2-\overline{\Omega}^2\partial^{ \sigma}\Omega^2\right)\,,\ee
where $\Delta\equiv \Omega^4+4\overline{\Omega}^4$. 
Substituting $C$ back into the action, we obtain the corresponding metric action
$$ {\cal{S}}=\int {\rm d}^4x\sqrt{-g}\left\{\frac{1}{2}\Omega^2R(g)+\frac{3}{4}\frac{(\nabla\Omega^2)^2}{\Omega^2}-\frac{3}{\Omega^2\Delta}\left(\Omega^2\nabla\overline{\Omega}^2-\overline{\Omega}^2\nabla\Omega^2\right)^2\right.$$
\be\left.-\frac{1}{4\gamma}(\Omega^2-1)^2-\frac{1}{4\delta}(\overline{\Omega}^2-\beta)^2\right\}\,.\ee

\subsection{The Einstein frame}

\begin{table}
\caption{List of different combinations of the curvature scalars in metric-affine  gravity (left column) and the equivalent metric theory (right column).  Note that the dynamical scalar field $\sigma$ is not present in all cases, and the viability of inflation varies.}
\begin{center}
\label{table1}
\begin{tabular}{c@{\quad}cl}
\hline
\multicolumn{1}{l}{\rule{0pt}{12pt}
                   metric-affine gravity}&\multicolumn{2}{l}{\,\,\,\,\,Metric gravity}\\[2pt]
\hline\rule{0pt}{12pt}
$\mathcal{R} $ &    GR & \\
  $\mathcal{R} + \tilde{\mathcal{R}}  $  &   GR& \\
$\mathcal{R} + \mathcal{R}^2 $  &   GR& \\
$\mathcal{R} + \tilde{\mathcal{R}}^2 $   &  $\sigma$, No inflation& \\
$\mathcal{R} + \tilde{\mathcal{R}} +\mathcal{R}^2  $ & $\sigma$, Inflation possible& \\
$\mathcal{R} + \tilde{\mathcal{R}} +\tilde{\mathcal{R}}^2  $ & $\sigma$, Inflation possible& \\
$\mathcal{R} + \tilde{\mathcal{R}}^2 +\mathcal{R}^2  $ & $\sigma$, No inflation& \\
$\mathcal{R} + \tilde{\mathcal{R}} +\mathcal{R}^2 +\tilde{\mathcal{R}}^2  $ & $\sigma$, Inflation possible& \\[2pt]
\hline
\end{tabular}
\end{center}
\end{table}
The Weyl rescaling $g_{ \mu\nu}=\Omega^{-2}\bar{g}_{ \mu\nu}$ takes us to the Einstein frame. Note that the Ricci scalar is rescaled as $R(g)=\Omega^2\bar{R}-6\Omega^3\Box\Omega^{-1}$. The action is
$$ {\cal{S}}=\int {\rm d}^4x\sqrt{-\bar{g}}\left\{\frac{1}{2}\bar{R}(\bar{g})-\frac{3}{\Omega^4\Delta}\left(\Omega^2\bar{\nabla}\overline{\Omega}^2-\overline{\Omega}^2\bar{\nabla}\Omega^2\right)^2\right.$$
\be\left.-\frac{1}{\Omega^4}\left(\frac{1}{\gamma}(\Omega^2-1)^2+\frac{1}{\delta}(\overline{\Omega}^2-\beta)^2\right)\right\}\,.\ee
Introducing the field $\sigma \equiv \overline{\Omega}^2/(2\Omega^2)\,,$
the scalar part of the Lagrangian becomes
\be {\cal{L}}=-\frac{12(\bar{\nabla}\sigma)^2}{(1+16\sigma^2)}-\frac{1}{4}\left(\,\frac{1}{\gamma}(\Omega^{-2}-1)^2+\frac{1}{\delta}(2\sigma-\beta\Omega^{-2})^2\,\right)\,.\ee
Variation with respect to the non-dynamical $\Omega^2$ gives
\be \frac{\delta{\cal{L}}}{\delta\Omega^2}=0\,\Longrightarrow\,\Omega^{-2}=\frac{\delta+2\beta\gamma\sigma}{\delta+\beta^2\gamma}\,\Longrightarrow\,{\cal{L}}=-\frac{12(\bar{\nabla}\sigma)^2}{(1+16\sigma^2)}-\frac{1}{4}\frac{(2\sigma-\beta)^2}{(\delta+\beta^2\gamma)}\,.\ee
The theory can be expressed in terms of a canonical scalar $s$ defined by $ \sigma=\frac{1}{4}\sinh(\sqrt{2/3}\, s)$
as
\be {\cal{L}}=-\frac{1}{2}(\nabla s)^2-\frac{1}{16}\frac{\left(\sinh(\sqrt{2/3}\, s)-2\beta\right)^2}{(\delta+\beta^2\gamma)}\,.\ee
At least one of $\gamma$ and $\delta$ has to be included in order to generate the additional pseudoscalar degree of freedom represented by $\sigma$. The inflationary behaviour of this model has been studied by G.~Pardisi and A.~Salvio in~\cite{Pradisi:2022nmh}. Note that the parameters $\gamma$ and $\delta$, associated with ${\cal{R}}^2$ and $\tilde{\cal{R}}^2$ , can only have a secondary role in a possible inflationary behaviour, which would be controlled by $\beta$. See also Table~\ref{table1} for more details.

\section{Coupling to a fundamental scalar}
We consider a scalar $\phi$ coupled to quadratic metric-affine gravity non-minimally. The action is
\be{\cal{S}}=\int\,{\rm d}^4x\sqrt{-g}\left\{\frac{1}{2}f(\phi){\cal{R}}+\frac{1}{2}h(\phi)\tilde{\cal{R}}\,+\,\frac{\gamma}{4}{\cal{R}}^2+\frac{\delta}{4}\tilde{\cal{R}}^2+{\cal{L}}_{\phi}\right\}\,,\ee
with
\be {\cal{L}}_{\phi}=-\frac{1}{2}g^{ \mu\nu}\partial_{ \mu}\phi\partial_{ \nu}\phi\,-V(\phi)\,.\ee
Introducing the auxiliaries $\chi$ and $\zeta$, we arrive at
\be {\cal{S}}=\int {\rm d}^4x\sqrt{-g}\left\{\frac{1}{2}\Omega^2{\cal{R}}+\frac{1}{2}\overline{\Omega}^2\tilde{\cal{R}}-\frac{1}{4}\left(\frac{1}{\gamma}(\Omega^2-f(\phi))^2+\frac{1}{\delta}(\overline{\Omega}^2-h(\phi))^2\right)+{\cal{L}}_{\phi}\right\}\,,\ee
with $ \Omega^2=\gamma\chi+f(\phi) $ and $\overline{\Omega}^2=\delta\zeta+h(\phi)\,.$
Rewriting the action in terms of the distortion and solving for it we arrive at the action of the corresponding metric theory in the Jordan frame. The Weyl rescaling $g_{ \mu\nu}\rightarrow \Omega^{-2}g_{ \mu\nu}$ takes the action into the Einstein frame and after introducing the auxiliary field $ \sigma=\overline{\Omega}^2/(2\Omega^2)$, we get the action in the form
\ba
{\cal{S}}=\int {\rm d}^4x\sqrt{-g} &&\left\{\frac{1}{2}R-\frac{12(\nabla\sigma)^2}{(1+16\sigma^2)}-\frac{1}{2}\frac{(\nabla\phi)^2}{\Omega^2}-\frac{\sigma^2}{\delta} -\frac{1}{4\gamma\Omega^4}\left(f(\phi)-\Omega^2\right)^2\right. \nonumber
\\ && \left.-\frac{h(\phi)}{4\delta\Omega^4}\left(h(\phi)-  4\sigma\Omega^2\right)-\frac{V(\phi)}{\Omega^4}\right\}\,.
\ea
Note that no kinetic term for $\Omega^2$ appears. Solving for it we obtain
\be \frac{\delta{\cal{S}}}{\delta\Omega^2}=0\,\Longrightarrow\,\,
\Omega^2=\frac{f(\phi)^2+4\gamma V(\phi)+\gamma h^2(\phi)/\delta}{f(\phi)+2\gamma\sigma h(\phi)/\delta-\gamma(\nabla\phi)^2}\,.\ee
Substituting $\Omega^2$ into the action we get it in the form
\be {\cal{S}}=\int {\rm d}^4x\sqrt{-g}\left\{\frac{1}{2}R-\frac{1}{2}K_{ \phi}(\phi,\sigma)(\nabla\phi)^2+\frac{1}{4}L_{ \phi}(\phi)(\nabla\phi)^4-\frac{1}{2}K_{ \sigma}(\sigma)(\nabla\sigma)^2-U(\phi,\sigma)\right\}\ee 
where
\ba
K_\phi(\phi,\sigma) &=& \frac{f(\phi)+2\gamma\sigma h(\phi)/\delta}{\gamma h(\phi)^2/\delta+f^2(\phi) +4\gamma V(\phi)}\,,\nonumber
\\
L_\phi(\phi) &=&  \frac{\gamma}{\gamma h(\phi)^2/\delta+f^2(\phi)+4\gamma V(\phi)}\,, \nonumber
\\
K_\sigma(\sigma) &=& \frac{24}{1+16\sigma^2}\,, 
\\
 V_{\rm eff}(\phi,\sigma) &=& \frac{V(\phi)}{f^2(\phi)+4\gamma V(\phi)}+\frac{1}{\delta}\left(\frac{f^2(\phi)+4\gamma V(\phi)}{\gamma h^2(\phi)/\delta+f^2(\phi)+4\gamma V(\phi)}\right)\left(\sigma-\sigma_0(\phi)\right)^2\,. \nonumber
\ea
where
\be \sigma_0(\phi)=\frac{h(\phi)f(\phi)/2}{f^2(\phi)+4\gamma V(\phi)}\,.\ee
Note that the potential is positive-definite with a minimum line along $\sigma=\sigma_0(\phi)$.
\begin{figure}[t!]  
\begin{center}
\includegraphics[scale=0.578]{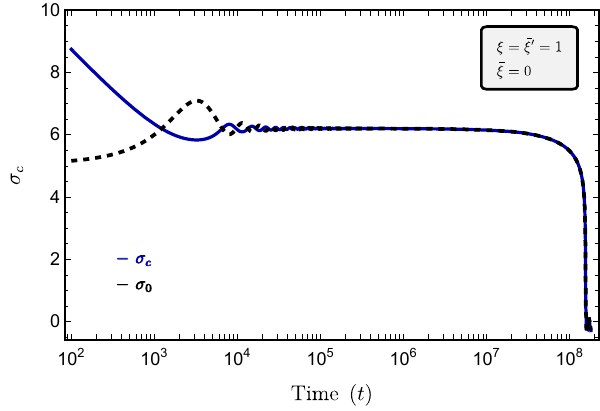}
\includegraphics[scale=0.432]{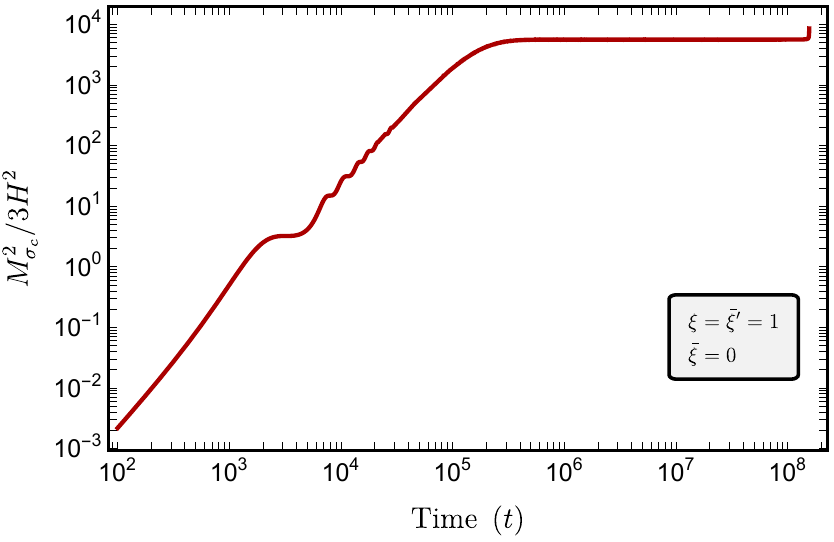}
\caption{{\textbf{Left:}} The evolution of the scalar field $\sigma_c$ and the minimum direction $\sigma_0(\phi)$. {\textbf{Right:}} The evolution of the ratio $M_{\sigma_c}^2(\phi)/(3H^2)$.  In both panels $\gamma=10^6$, $\xi=\bar{\xi}'=1$ and $\bar{\xi}=0$.  }
\label{fig:1}
\end{center}
\end{figure}
Along the minimum line the potential is just $U_0(\phi)=\frac{V(\phi)}{f^2(\phi)+4\gamma V(\phi)}$. However the kinetic term of $\phi$ is modified by $-\frac{1}{2}K_{ \sigma}(\sigma_0(\phi))(\nabla\sigma_0(\phi))^2$. The effective Lagrangian is
\be {\cal{L}}_{eff}=-\frac{1}{2}\overline{K}(\phi)(\nabla{\phi})^2+\frac{1}{4}L(\phi)(\nabla\phi)^4-U_0(\phi)\,,\ee  
where $\overline{K}(\phi)$ is the modified kinetic function~\cite{Gialamas:2022xtt}.

\begin{figure}[t!]  
\begin{center}
\includegraphics[scale=0.65]{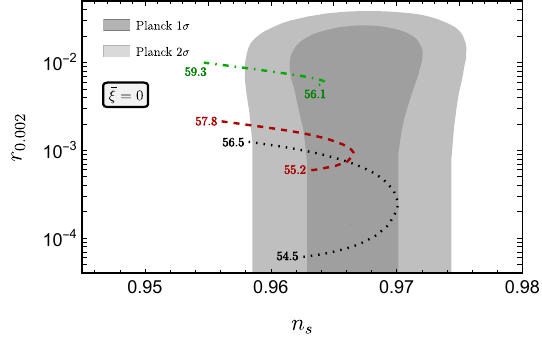}
\includegraphics[scale=0.65]{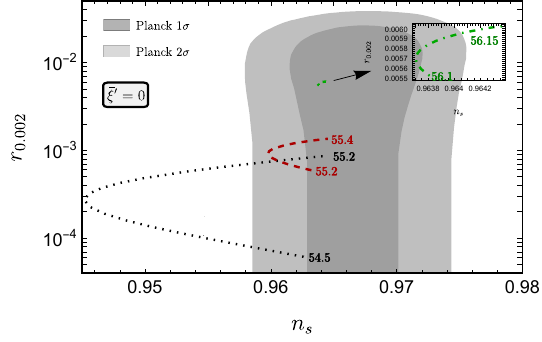}
\caption{
Predictions of the model in the $n_s-r$ plane. {\textbf{Left:}} The case $\bar{\xi}=0$, with running $\bar{\xi}'$. {\textbf{Right:}} The case $\bar{\xi}'=0$, with running $\bar{\xi}$. In both panels, $\xi=0.1$ (green), $\xi=1$ (red) and $\xi=10$ (black). The small numbers at the edges of the curves indicate the number of $e-$folds. }
\label{fig:2}
\end{center}
\end{figure}
\section{Inflation}
In what follows we shall adopt the following leading terms of $f(\phi)$ and $h(\phi)$, namely
\be f(\phi)=1+\xi\phi^2,\,\,\,\,\,\,\,h(\phi)=\bar{\xi}\phi+\bar{\xi}'\phi^3\,.\ee 
Note that $h(\phi)$ is chosen this way to counteract the parity-odd coupling $h(\phi)\tilde{\cal{R}}$. We also replace $\sigma$ with the canonical field $\sigma_c=2\sqrt{6}\int\frac{d\sigma}{\sqrt{1+16\sigma^2}}$.
In a FRW background the equations of motion read
\ba
&&(K_{ \phi}+3L_{ \phi}\dot{\phi}^2)\ddot{\phi}+3H(K_{ \phi}+L_{ \phi}\dot{\phi}^2)\dot{\phi} + \dot{\phi} \dot{\sigma}_{c} \frac{\partial K_{ \phi}}{\partial\sigma_{c}} +\left(\frac{1}{2}\frac{\partial K_{ \phi}}{\partial\phi}+\frac{3}{4}\frac{\partial L_{ \phi}}{\partial\phi}\dot{\phi}^2\right)\dot{\phi}^2+\frac{\partial V_{eff}}{\partial\phi}=0\,, \nonumber
\\&&\ddot{\sigma}_{c}+3H\dot{\sigma}_{c}-\frac{1}{2}\frac{\partial K_{ \phi}}{\partial\sigma_{c}}\dot{\phi}^2+\frac{\partial V_{eff}}{\partial\sigma_{c}}=0\,, \quad H^2=\frac{1}{3}\rho\,, \quad \dot{H}=-\frac{1}{2}\left(\rho\,+\,p\right)\,, \nonumber
\ea
where $\rho$ and $p$ are the total energy density and pressure respectively. Starting with large initial values for the fields, the system, in a relatively short time, drops into the valley defined by the minimum line $\sigma=\sigma_0(\phi)$ and its further evolution is described by a single field effective Lagrangian. This can be easily seen from the left panel of Fig.~\ref{fig:1}. In the right panel of Fig.~\ref{fig:1} is displayed the relevant mass of the $\sigma_c-$ field, $M_{\sigma_c}^2(\phi)$, in terms of the inflationary Hubble scale $H$. 
Following a numerical computation, it becomes evident that at times $ t > 10^5$, the ratio $M_{\sigma_c}^2(\phi)/H^2 \sim 10^4$ (in the benchmark scenario). This finding provides additional support to the inference that the model is equivalent to a single-field inflationary model.
In Figure~\ref{fig:2}, we present the model's predictions using pivot scales of $0.05\, {\rm Mpc^{-1}}$ for the spectral index $n_s$ and $0.002\, {\rm Mpc^{-1}}$ for the tensor-to-scalar ratio $r$. The regions filled with shading correspond to parameter ranges allowed at $68\%$ and $95\%$ confidence levels, as derived from~\cite{Akrami:2018odb,BICEP:2021xfz}. In the left (right) panel is displayed the case $\bar{\xi}=0$ ($\bar{\xi}'=0$).

\section{Conclusions}
Our investigation encompassed a broad quadratic metric-affine theory incorporating an additional dynamic degree of freedom, coupled non-minimally with a scalar field. Our exploration delved into the realm of two-field inflation, revealing its effective simplification to a single-field model. The model exhibited a potential very similar to the one in the Palatini-${\cal{R}}^2$ models. Remarkably, the inflationary predictions originating from this model align with the latest observational constraints over a wide range of parameters. Moreover, this framework presents the possibility of elevating the tensor-to-scalar ratio.
\section*{Acknowledgments}

The work of IDG was supported by the Estonian Research Council grants SJD18, MOB3JD1202, RVTT3,  RVTT7, and by the CoE program TK202 ``Fundamental Universe''.

\bibliography{proc_bib}{}

\end{document}